\documentclass[prb,twocolumn]{revtex4}
\usepackage{epsfig,graphicx}

\begin{document}

\title{Linear in-plane magnetoconductance and spin susceptibility
of a 2D electron gas on a vicinal silicon surface }
\author{Y.Y. Proskuryakov$^{1,*}$, Z.D. Kvon$^2$, A.K. Savchenko$^1$}

\address {$^1$ School of Physics, University of Exeter, Stocker Road, Exeter, EX4 4QL, U.K.\\
\ $^2$ Institute of Semiconductor Physics, Novosibirsk, 630090,
Russia}

\begin{abstract}
In this work we have studied the parallel magnetoresistance of a 2DEG near a vicinal silicon surface.  An unusual, linear
magnetoconductance is observed in the fields up to $B = 15$ T, which we explain by the effect of spin polarization on
impurity scattering. This linear magnetoresistance shows strong anomalies near the boundaries of the minigap in the
electron spectrum of the vicinal system.
\end{abstract}


\maketitle

Over the last few years a number of reports have appeared on observations of anomalous positive magnetoresistance
(negative magnetoconductance, MC) in high-mobility 2D electron and hole gases in the field parallel to the 2D plane
\cite{0.5}. Such interest is fuelled by the fact that this negative MC is directly related to other unusual properties of
high-mobility systems, such as `metallic' behaviour and the transition from `metal' to `insulator' \cite{0.5}. However,
until now all studies of in-plane MC were devoted to the range of low electron densities, $N_s$, near the
`metal-insulator' transition. In the case of a 2D electron gas (2DEG) in Si-MOSFET structures the densities studied were
below $10^{12}$ cm$^{-2}$.

The focus of our study is the in-plane MC in a \textit{vicinal} Si-MOSFET at \textit{high electron densities}, $N_s >
10^{12}$ cm$^{-2}$. It is well known \cite{0.7,1,6,6.5} that this system has a superlattice potential on the Si surface,
which results in a minigap in the energy spectrum. Previously, the study of the slow electron diffraction by atomically
pure vicinal Si surfaces cut at small angles $\theta$ to (100) plane (as shown in Fig. 1a) revealed ordered steps, which
do not disappear even when specimens are heated up to 1100$^o$C, even in the presence of hydrogen or oxygen \cite{6.6}.
The size of these steps was in agreement with theoretical predictions for perfect high-index surfaces \cite{6}. This
periodic structure is considered to be the reason for the appearance of the superlattice potential \cite{1,6}.

The minigap in the energy spectrum has an inter-valley character \cite{6}, which implies a strong inter-valley interaction
when the electron energy approaches the minigap. This also leads to the appearance of a logarithmic divergence in the
density of states $D(E)$ at energies close to the lower edge of the minigap, Fig. 1(b).  $D(E)$ differs from zero within
the minigap because the superlattice is one-dimensional and does not produce a full gap in the spectrum. The discontinuity
in $D(E)$ also appears at the upper edge of the minigap. When the smearing $\Gamma$ is less than the width of the minigap
$\Delta$, transport coefficients exhibit singularities as the Fermi energy $E_F$ passes through the boundaries of the
minigap \cite{1,6,10}.

Such systems with a minigap are of interest because they can realise two different situations: with an isotropic Fermi
surface (FS), and an anisotropic one, dependent on the position of the Fermi level. These two cases correspond to
different ranges of electron densities, where the properties of the 2DEG significantly differ: $N_s < N_{\Delta }^{'}$
(isotropic FS) and $N_s > N_{\Delta }^{'}$ (anisotropic FS), where $N_{\Delta }^{'} = \pi (0.15/L)^2 $ is the density
corresponding to the onset of the superlattice minigap. (Here $L = a/(2\sin (\theta))$ is the superlattice period,
$\theta$ is the angle between the vicinal and (100) silicon surface, $a = 5.43 \,\textrm{\AA}$ is the lattice constant of
Si.) The properties of the 2DEG in the first range are identical to those of an ordinary (100) Si-MOSFET, while in the
second range the properties of the 2DEG become strongly modified \cite{1,6}.

We experimentally investigate the MC in both ranges of $N_s$. A linear magnetoconductance is observed in magnetic field
parallel to the plane of the 2DEG, in a surprisingly large range of fields up to 15 T. To our knowledge, this is the first
observation of such an effect. We describe the linear decrease of the conductance with $B_{||}$ in terms of the screening
model \cite{7}, where impurity scattering gets changed when the 2DEG is spin-polarised by parallel magnetic field. The
performed analysis suggests that the slope of the linear MC in the first range of $N_s$ (isotropic Fermi surface) is in
agreement with theoretical expectation for the scattering dominated by impurity potential. In the second range of $N_s$ a
significant decrease of the magnitude of the MC is observed at the boundaries of the minigap. This can be attributed to
either the modification of the transport properties of the electrons near the points of topological transitions of the
Fermi surface \cite{8,10}, or suppression of the spin susceptibility.

The studied samples are Hall-bar Si-MOSFETs fabricated on a vicinal Si(17,2,2) surface, tilted from (100) by an angle
$\theta =9^\circ $ 40' around the direction [011]. The superperiod in this case is $L=16.2~\AA$, which corresponds to
$N_{\Delta }^{'}\simeq2.6\times10^{12}$ cm$^{-2}$. As this surface is just slightly different from the (811) surface at
$\theta=10^o$, one can expect qualitatively the same dispersion relation and other properties as those for Si(n11)
described in detail in Ref.4. We study the behaviour of the 2DEG for both orientations of the Hall bar -- along and
perpendicular to the superlattice axis. In the latter case, shown in Fig. 1(a), the peak mobility of the 2DEG is $\sim
25000$ cm$^2/$Vs. Our measurements have been carried out in magnetic fields up to 15 T at $T \simeq 50$ mK.  The
experiments have been performed at large electron densities: $N_s$ from $1\times10^{12}$ to $5\times10^{12}$ cm$^{-2}$.

In magnetic field parallel to the plane of the 2DEG we have observed negative MC in the entire range of $N_s$. An example
is shown in Fig. 1(c, d) for the Hall bar oriented parallel to the superlattice axis.
\begin{figure}[t]
\begin{center}
\includegraphics*[totalheight=3.0in]{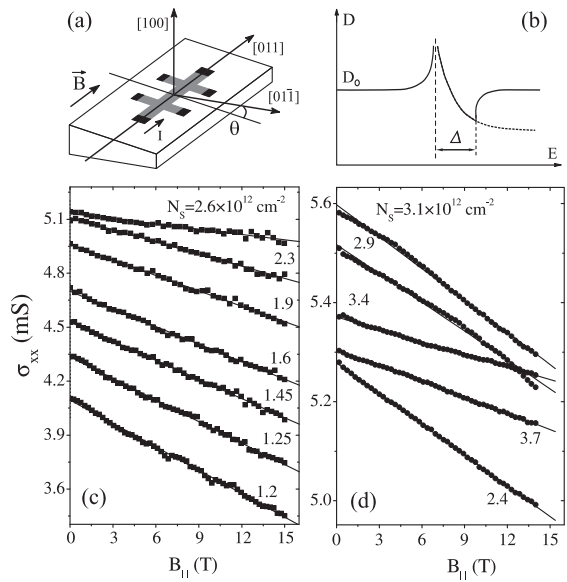}
\caption{(a) A schematic diagram showing the geometry of the studied sample in the case of the Hall-bar oriented
perpendicular to the super-lattice axis. (b) Density of states of 2DEG in a vicinal sample with super-lattice potential at
the surface. $D_0$ is unperturbed density of states (as in (100)Si MOSFETs), and $\Delta$ is the minigap width. (c-d)
Longitudinal conductivity versus in-plane magnetic field for different electron densities (symbols), with linear fits
(solid lines).}
\end{center}
\end{figure}
In the whole range of fields the conductivity decreases linearly with $B_{||}$. The slope of this linear dependence
decreases monotonically with increasing density up to $N_s \sim 2.5\times10^{12}$ cm$^{-2}$, which is close to
$N_{\Delta}^{'}$, Fig. 1(c). However, at larger densities the slope starts changing nonmonotonically, as seen in Fig.
1(d). It is important to emphasise that MC remains linear in the entire density range.

In Fig. 2 we show the conductivity measured as a function of electron density at different magnetic fields from 0 to 15 T,
for both Hall-bar orientations. In zero magnetic field the so called ``$\Omega$"- and ``W"- features are clearly seen in
$\sigma_{xx}(N_s)$ for the two orientations, respectively (the names reflect the shapes of the two dependences in the
minigap). These well known features were previously observed in vicinal systems \cite{1,6}. They originate from the
superlattice structure of the vicinal surface and indicate the very good quality of our samples.
\begin{figure}[b]
\begin{center}
\includegraphics*[totalheight=3.5in]{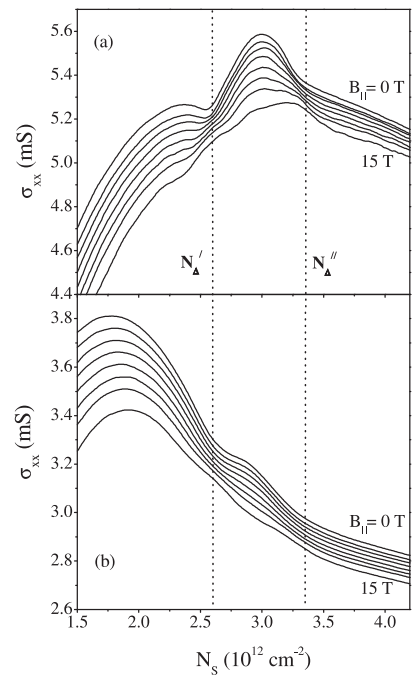}
\caption{Density dependence of the conductivity at different in-plane magnetic fields: $B_{||}=0$, 2, 4, 6, 8, 10, 12, 15
T; (a) for the Hall-bar oriented perpendicular to the supperlattice axis, (b) along the supperlattice axis. The dotted
lines mark the boundaries of the superlattice minigap at $N_{\Delta}^{'}$ and $N_{\Delta}^{''}$}.
\end{center}
\end{figure}

One can notice that the effect of magnetic field on the conductivity is minimal in the characteristic points corresponding
to the lower and upper boundaries of the superlattice minigap, $N_{\Delta}^{'}$ and $N_{\Delta}^{''}$, marked by the
vertical dotted lines in Fig. 2 (a,b). This is seen for both Hall-bar orientations and corresponds to the non-monotonic
dependence of the linear MC on electron density in Fig. 1(b). Also, a transformation of the ``W" and ``$\Omega$" features
is seen in Fig. 2 (a,b) -- these features seem to be significantly weakened when a strong in-plane magnetic field is
applied.

To analyse the linear MC let us first consider the range $N_s < N_s^{\Delta}$, where, as mentioned above, the properties
of a 2DEG near the vicinal surface and the (100) surface are expected to be equivalent. Also, it is well established that
scattering in the low density regime is dominated by impurity scattering rather than by interface roughness \cite{1}. In
this region of $N_s$ the slope of the linear MC decreases monotonically with increasing density as it has been shown
above.

Among several models of negative parallel-field MC \cite{11}, there are two which indeed predict linear MC: one is the
interaction theory by Zala, Narozhny and Aleiner \cite{7.5}, and the other one is the screening model by Dolgopolov and
Gold \cite{7}. We start our consideration from the interaction theory, which addresses the effect of in-plane magnetic
field on quantum correction to conductivity due to electron-electron ($e-e$) interactions in 2D disordered systems. This
theory predicts linear MC in relatively strong magnetic fields and in the ballistic regime, when the parameter
$k_BT\tau/\hbar\gg 1$. According to Ref. [10] the strong field criterium ($g^*\mu_B B_{||}/2k_BT \gg 1$, where $g^{*}$ -
Lande g-factor, $\mu _B$ - the Bohr magneton) is realised in our experiment already at $B_{||}>0.3$ T. However the values
of the parameter $k_BT\tau/\hbar$ range from 0.003 to 0.015, implying that the system is in the diffusive regime,
$k_BT\tau/\hbar\ll 1$. The estimation of MC caused by electron interactions in this regime \cite{7.5} shows that effect of
$e-e$ interactions is negligible, as the magnitude of MC (proportional to $\ln (B)$) raises to only about $1\%$ at $B_{||}
\sim 10$ T.

In the screening model \cite{7,7.2}, developed for the system with isotropic Fermi surface and at $T=0$, the MC effect
originates from the change in the screening of scattering potential caused by the difference between Fermi momenta
$k_{F-}$ and $k_{F+}$ for the two spin-split subbands. It predicts the change of the Drude conductivity with the following
negative MC in the low-density case, $q_s\gg 2k_F$, where $q_s$ is the screening wave number. At small magnetic fields
$B_{||}$ such that $B_{||}<0.2B_S$, where $B_S=2E_F/g^{*}\mu _B$ is the field of the full spin polarisation, this MC is
linear in the case of short-range impurity scattering:
\begin{equation}
\frac{\sigma (B_{||}<0.2B_S)}{\sigma (B_{||}=0)}\simeq1- \alpha \frac{B_{||}}{B_S}, \label{eq1}
\end{equation}
where the coefficient $\alpha$ varies in the range from 0.87 to 0.68 for the electron densities from $N_s=1\times10^{12}$
to $5\times10^{12}$ cm$^{-2}$, Ref. [8-9]. As the latter is the range of $N_s$ of our experiment, we take the average
value of $\alpha=0.78$ for the approximate analysis.

In our case $q_s/2k_F \sim \, 2-4$ which justifies the approximation of the low density. To justify the applicability of
the theory developed for $T=0$, we only analyse quantitatively the data at the lowest temperature $T=50$ mK. Simple
estimation using standard parameters for (100) Si MOSFETs ($g^*\sim 2.5$, $m^*=0.19 m_e$) gives then $10-15\%$
magnetoconductivity in magnetic field $B \sim 15$ T, which is close to what is seen in Fig. 1(a). The screening theory
analyses the effect of magnetic field in both cases of impurity and roughness scattering provided $B_{||}=B_S$, Ref. [9].
However, the expression for the small field MC has been obtained only for the impurity scattering, Eq. (\ref{eq1}).

As a first approximation, we analyse the slope of the linear dependence in the whole range of electron densities using Eq.
(\ref{eq1}) and neglecting the influence of interface-roughness scattering. It follows from the above definition of $B_S$
that $B_S^{-1}\propto g^*m ^*$ and hence it is proportional to the spin susceptibility of the electron gas $\chi$.
(Indeed, for a 2DEG in a Si MOSFET $E_F=\pi\hbar^2N_s/2m^*$ and $\chi=2\mu_B^2 g^* m^*/\pi \hbar^2$ \cite{1}). The
electron density is known from the measurements of Shubnikov--de Haas (SdH) oscillations. From the slope of the linear MC
the product $g^* m^*$ is then found and normalised by $g_0 m_b$, where $m_b=0.19m_e$ is bulk electron mass ($m_e$ is the
free electron mass) and $g_0=2.0$ is the Lande g-factor in bulk silicon. The data for the both Hall-bar orientations are
plotted in Fig. 3(a, b).
\begin{figure}[hbtp]
\begin{center}
\includegraphics*[totalheight=3.5in]{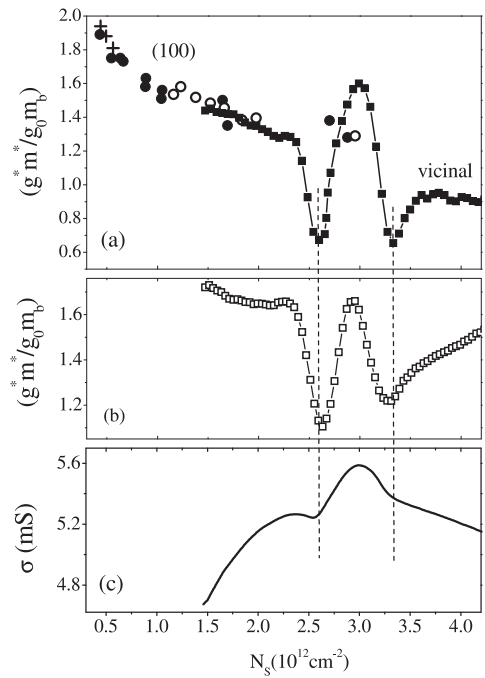}
\caption{(a, b) Density dependence of the product $g^*m^*$ (proportional to spin susceptibility $\chi$) normalised by the
product of the bulk parameters $g_0m_b$: (a) for the Hall bar oriented perpendicularly to the superlattice axis, (b) along
this axis. Data for (100)Si MOSFETs: ($\circ$) - from Ref.[13], ($\bullet$) - from Ref.[14], ($+$) - from Ref.[15]. (c)
Density dependence of the conductivity at $B=0$ for the Hall-bar in case (a) (the same data as in Fig. 2 (a).}
\end{center}
\end{figure}
In Fig. 3(a) we also show the $g^*m^*$ results obtained by three different experimental groups using the analysis of SdH
oscillations in ordinary (100)Si MOSFETs \cite{13,13.1,13.2}. At lower densities our results show a monotonic increase of
spin susceptibility with decreasing $N_s$. This trend at $N_s\leq 2.3\times10^{12}$ cm$^{-2}$ as well as the magnitude of
$g^*m^*$ agree well with the previous results. The above range is exactly the first range of $N_s$ we referred to before
($N_s < N_{\Delta }^{'}$), where the Fermi surface of the vicinal sample is isotropic, as it is in the case of (100)Si,
and where the screening theory is valid. This result also indicates that in this range of $N_s$ the effect of the
interface-roughness scattering is negligible.

A strong deviation from the monotonic behaviour of spin susceptibility is observed at larger densities ($N_s> N_{\Delta
}^{'}$) if the analysis is carried out using the same approach as above.  A drop of $ g^*m ^*$ by more than a factor of
two is seen at two different $N_s$ for the Hall-bar oriented perpendicularly to the superlattice axis, Fig. 3(a). The two
minima also appear for the other Hall-bar orientation (along the superlattice axis), Fig. 3(b), although they are less
pronounced, presumably because of the smaller mobility of this sample (the difference in the conductivities is seen in
Fig. 2). It is clearly seen from the comparison to Fig. 3(c) that these minima coincide with the dips of the ``$\Omega$"
and ``W" features in $\sigma(N_S)$ -- the points where the Fermi level crosses the boundaries of the superlattice minigap.
(The pronounced decrease of MC at these points is also seen in Fig. 2 a, b.) It is important to notice that the changes of
the conductivity with $N_s$ at the minigap boundaries do not exceed 10\%, Fig. 2, while a much more drastic decrease is
seen in the density dependence of the spin susceptibility.

We have to note, however, that the screening theory is not expected to be valid in the range of $N_s > N_{\Delta }^{'}$.
At the same time, it is seen in Fig. 3(a) that in the middle of the minigap the obtained values of $ g^*m ^*$ are close to
those obtained for (100)Si using SdH oscillations. This points out that the anomalies in the MC arise only at the
boundaries of the minigap.

Let us discuss possible reasons for the dramatic decrease of the MC at the minigap boundaries. A typical dispersion
relation of the vicinal system in Si is shown in Fig. 4(a). Different configurations of the Fermi surface are shown in
Fig. 4(b-e) for four positions of the Fermi level. It is seen on the plot that the boundaries of the minigap in the
spectrum correspond to the topological transitions of the Fermi surface: firstly the two isotropic and independent FSs
coalesce (Fig. 4c), and then they form two inclosed surfaces (Fig. 4e).
\begin{figure}[t]
\begin{center}
\includegraphics*[totalheight=4.0in]{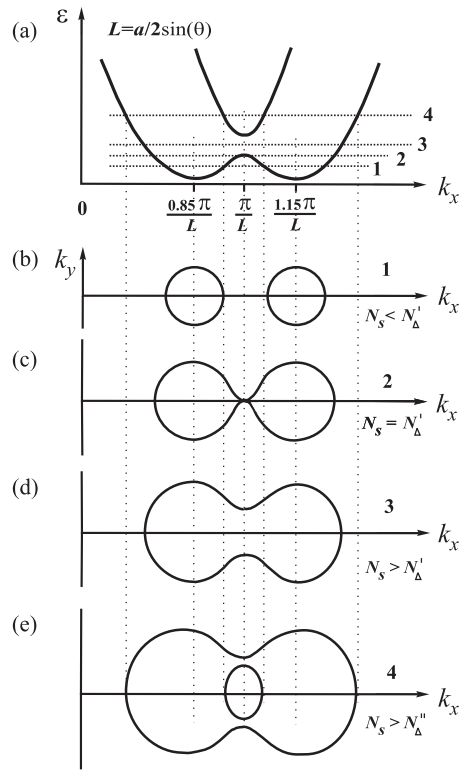}
\caption{(a) Typical dispersion relation for a 2DEG in a Si vicinal system \cite{6}. (b-e) Fermi surfaces at different
positions of the Fermi level $E_F$ marked by horizontal dotted lines in (a).}
\end{center}
\end{figure}
It is known that the conductivity and thermopower of the electron system is strongly modified at the points of the
topological transitions ($N_s=N_{\Delta }^{'},~N_{\Delta }^{''}$) \cite{6,10}. In our case we see a strong decrease of the
linear MC at these points.

The exact theory of MC for complicated FS is not developed. However, if one assumes that the screening theory \cite{7,7.2}
can still be applied at the boundaries of the minigap, our results indicate that the spin susceptibility sharply decreases
in the points of the topological transitions. Continuing this logic, it is important to note that previous investigations
\cite{8} have shown that the density of states at the minigap boundaries changes only within $10\%$, which would
correspond to a small change in the effective mass. This implies that it is mainly the g-factor that is affected by the
topological transitions in the 2DEG spectrum of the vicinal system.

An increase of the g-factor, caused by enhanced exchange interaction, is well known \cite{12}. In contrast, in our case a
significant decrease of $g^*$ (by a factor of two) is observed. This effect could possibly be caused by the strong
inter-valley interaction in the vicinal structure (giving rise also to the ``$\Omega$" and ``W" features in the
conductivity \cite{1}). We have to say, however, that the above speculations are based on the assumption that the
screening theory \cite{7,7.2} is applicable within the minigap. An appropriate theory for such a case does not exist at
the moment, and we hope that our experiments will stimulate its development.

This work was supported by Royal Society, RFBR (grant 02-02-16516), INTAS (project 01-0014), Russian Ministry of Education
(program ``Integration") and RAS (program ``Low dimensional quantum structures"), EPSRC and ORS scheme.

* Current address: Physics Department, Royal Holloway, University of London, Egham, Surrey, TW20 0EX, U.K.

\end{document}